\documentstyle[preprint,aps]{revtex}

\tightenlines
\begin{document}
%\twocolumn
\draft %\draft command makes pacs numbers print
%\preprint{1. version}
\title{Comment on "Quantum Theory of Dispersive Electromagnetic Fields"}
\author{Gediminas Juzeli\={u}nas}
\address{Institute of Theoretical Physics and Astronomy,\\
A. Go\v{s}tauto 12, \\
2600, Vilnius, Lithuania }
\date{\today }
\maketitle
\begin{abstract}
Recently Drummond and Hillery \lbrack Phys. Rev. A {\bf 59}, 691
(1999)\rbrack\ presented a quantum theory of dispersion based on the
analysis of a coupled system of the electromagnetic field and atoms in the
multipolar QED formulation. The theory has led to the explicit
mode-expansions for various field-operators in a homogeneous medium
characterized by an arbitrary number of resonant transitions with different
frequencies. In this Comment, we drawn attention to a similar multipolar
study by Juzeli\={u}nas \lbrack Phys. Rev. A {\bf 53}, 3543 (1996); A {\bf %
55}, 929 (1997)\rbrack\ on the field quantization in a discrete molecular
(or atomic) medium. A comparative analysis of the two approaches is carried
out highlighting both common and distinctive features.
\end{abstract}
\pacs{PACS 42.50.-p, 32.80.-t, 71.36.+c, 12.20.Ds}
\narrowtext 

Recently Drummond and Hillery \cite{drum:99} have presented a quantum theory
of dispersive electromagnetic modes in a medium characterized by an
arbitrary number of resonant transitions with different frequencies. The
radiation field and the matter was assumed to constitute a single dynamical
system. In the homogeneous (plane wave) case, explicit expansions have been
obtained for the field-operators in terms of the operators for creation and
annihilation of polaritons (elementary excitations of a combined system
containing the radiation and the matter).Yet, the authors of the paper \cite
{drum:99} have overlooked a closely related study \cite{juz:96,juz:97} that
also provides explicit mode-expansions for various quantized fields, such as
the macroscopic displacement field, given by (in the Heisenberg picture) 
\cite{juz:96}:

\begin{equation}
\overline{{\bf d}}^{\bot }\left( {\bf r},t\right) =\sum\limits_{{\bf k},m}%
\overline{{\bf d}}_{{\bf k},m}^{\bot }\left( {\bf r},t\right) ;  \label{d}
\end{equation}
with 
\begin{equation}
\overline{{\bf d}}_{{\bf k},m}^{\bot }\left( {\bf r},t\right)
=i\sum\limits_{\lambda =1}^{2}\left( \frac{\varepsilon _{0}\hbar
kv_{g}^{\left( m\right) }}{2V_{0}}\right) ^{1/2}n_{{}}^{\left( m\right) }%
{\bf e}_{{}}^{\left( \lambda \right) }\left( {\bf k}\right) \left[
e^{i\left( {\bf k.r-}\omega _{k}^{\left( m\right) }t\right) }P_{{\bf k}%
,m,\lambda }^{{}}-e^{-i\left( {\bf k.r-}\omega _{k}^{\left( m\right)
}t\right) }P_{{\bf k},m,\lambda }^{+}\right]  \label{d-km}
\end{equation}
where $P_{k,m,\lambda }^{+}$ ($P_{k,m,\lambda }^{{}}$) is the Bose operator
for creation (annihilation) of a polariton characterized by a wave-vector $k$%
, a polarization index $\lambda $, and also an extra index $m$ $=1,2,...,M+1$
that labels branches of polariton dispersion, $M$ being a\ number of
excitation frequencies accommodated by each molecule forming the medium.
Here also ${\bf e}_{{}}^{\left( \lambda \right) }\left( {\bf k}\right) $ is
a unit polarization vector, $V_{0}$ is the quantization volume, $\omega
_{k}^{\left( m\right) }=ck/n_{{}}^{\left( m\right) }$ is the polariton
frequency, $n_{{}}^{\left( m\right) }\equiv n\left( \omega _{k}^{\left(
m\right) }\right) $ is the refractive index (calculated at $\omega
_{k}^{\left( m\right) }$), and $v_{g}^{\left( m\right) }=d\omega
_{k}^{\left( m\right) }/dk$ is the branch-dependent group velocity.

In what follows we compare the formalism by Drummond and Hillery \cite
{drum:99} to that by Juzeli\={u}nas \cite{juz:96,juz:97,juz:95} highlighting
common and distinctive features of the two approaches. Both studies consider
a similar coupled system of the radiation field and the matter, exploiting
the same multipolar formulation of Quantum Electrodynamics (QED). (Yet,
different techniques have been employed to represent the operators of
interest in terms of the normal polariton modes.) Moreover, in either
approach an arbitrary number of transition frequencies (of electronic or
vibrational origin) has been included for the material medium. As a result,
the above mode-expansion (\ref{d})-(\ref{d-km}) reproduces the same
functional dependence on the group velocity and the refractive index as the
corresponding expansion for the macroscopic displacement field in the
one-dimensional case, given by Eq.(5.17) of ref.\cite{drum:99}. Such a
functional dependence is also consistent with earlier narrow-band
Langrangian approach by Drummond \cite{drum:90}. Furthermore, the
mode-expansion (\ref{d})-(\ref{d-km}) is equivalent to the three-dimensional
displacement operator derived by Drummond and Hillery \cite{drum:99,footn:1}%
, as long as the spatial dispersion is neglected in the corresponding Eq.
(8.24) of ref.\cite{drum:99}. The same holds for other field-operators, such
as the operator for the transverse electric field whose mode-components are
related to Eq.(\ref{d-km}) via a relationship of the classical type \cite
{juz:96}: $\overline{{\bf d}}_{{\bf k},m}^{\bot }\left( {\bf r}\right)
=\varepsilon _{0}\varepsilon _{r}^{\left( m\right) }\overline{{\bf e}}_{{\bf %
k},m}^{\bot }\left( {\bf r}\right) $, the emerging relative dielectric
permittivity $\varepsilon _{r}^{\left( m\right) }\equiv \left(
n_{{}}^{\left( m\right) }\right) ^{2}$ being a branch-dependent quantity.
This is also in agreement with the ref. \cite{drum:99}.

Both studies take special care in making sure that the (equal-time)
commutation relationships preserve between various operators in their
diagonal (polariton) representation. The study \cite{juz:96,juz:97} has
checked the commutation relationships involving the field-operators $%
\overline{{\bf d}}^{\bot }$, ${\bf e}^{\bot }$, $\overline{{\bf a}}^{\bot }$%
, $\overline{{\bf p}}^{\bot }$ and $\overline{{\bf h}}^{\bot }$. Their
correctness appears to be ensured by the following equalities \cite
{juz:96,juz:97,footn:2}: 
\begin{equation}
\begin{array}{lll}
\sum\limits_{m}v_{g}^{\left( m\right) }n_{{}}^{\left( m\right) }=c & \text{%
and} & \sum\limits_{m}v_{g}^{\left( m\right) }/n_{{}}^{\left( m\right) }=c
\end{array}
\label{rel}
\end{equation}
Drummond and Hillery \cite{drum:99} have also exploited one of the these
equalities in analyzing the commutation relationships. In addition, a few
more complex equalities have been presented when dealing with the material
operators.

A distinctive feature of the approach by Drummond and Hillery \cite{drum:99}
is inclusion of spatial dispersion, i.e. dependence of the dielectric
permittivity not only on the frequency, but also the wave-vector \cite
{agran:84}. This does not alter the form of the above relationships, as well
as the mode-expansions of other macroscopic field-operators presented
earlier \cite{juz:96}, yet the meaning of group velocity is to be modified 
\cite{drum:99}. It is noteworthy that the spatial dispersion has been
included in the 'effective mass' approximation through an extra differential
term featured in Eqs.(6.2) and (6.8) of ref. \cite{drum:99}: The term
represents the coupling between infinitely close dipole-oscillators
comprising the continuous dielectric medium. One might argue that such an
approach is not fully consistent with the spirit of multipolar QED
formulation in which there is no direct coupling between the
dipole-oscillators \cite{multipol}. However, this is perhaps the only way to
include the spatial dispersion within the continuous model of the dielectric
considered.

On the other hand, the approach \cite{juz:96,juz:97,juz:95} assumes the
matter to be discrete, the constituent molecules (dipole-oscillators)
forming a cubic lattice. Here the effects of spatial dispersion (as well as
other intermolecular coupling) are contained in the initial multipolar
Hamiltonian for the radiation field coupled to the discrete medium, all
interatomic coupling being mediated exclusively via the transverse virtual
photons \cite{multipol}. As a result, the spatial dispersion is implicit in
the general analysis of ref. \cite{juz:96} up to Eq.(3.40). The effect has
been omitted in the subsequent long wave-length approximation made in
Eq.(4.1) of ref. \cite{juz:96} followed by the explicit results. In fact,
the spatial dispersion does play a minor role for the optical (photon-like)
modes characterized by small wave-vectors. One can recover the spatial
dispersion in a relatively\ straightforward manner using the discrete model 
\cite{juz:96}, however this is out of the scope of the present Comment. Note
that in contrast to the continuous approach \cite{drum:99}, the spatial
dispersion is characterized exclusively by the microscopic parameters of a
discrete system, there being no need to introduce an extra parameter
describing the effect.

Using a discrete approach, one can also recover the local field effects
from first principles. In doing this, the theory \cite{juz:96,juz:97,juz:95}
treats systematically the Umklapp processes playing an important role in the
multipolar QED formulation. As a result, the required local-field
corrections emerge intrinsically in the refractive index and the group
velocity entering the mode-expansions of the field-operators \cite
{juz:96,juz:97,juz:95}. Furthermore, the discrete approach allows us to
consider not only the macroscopic operators \cite{juz:96}, but also the
operators for the local \cite{juz:96,juz:97,juz:95} and microscopic \cite
{juz:97} fields. For instance, the mode-components of the local displacement
operator are related to those for the macroscopic displacement 
operator as \cite{juz:96}: 
${\bf d}_{{\bf k},m}^{\bot }\left( {\bf r}_{\zeta }\right) =\left(
\varepsilon _{r}^{\left( m\right) }\right) ^{-1}\left[ \left( \varepsilon
_{r}^{\left( m\right) }+2\right) /3\right] \overline{{\bf d}}_{{\bf k}%
,m}^{\bot }\left( {\bf r}_{\zeta }\right) $. This appears to be very helpful
for the analysis of various molecular-radiation processes in dielectric
media \cite{juz:96,juz:95}, such as the spontaneous emission.

Finally, Drummond and Hillery \cite{drum:99} have pointed out that the
solution to the eigenvalue equation $\omega ^{2}=c^{2}k^{2}/n^{2}$ given by
Eq.(7.5), 'is unique for any given modal frequency, but has forbidden
regions which indicate a resonance, or absorption, band'. This would be
absolutely true for the spatially non-dispersive media, as illustrated in
fig.2 of ref.\cite{juz:96}. However Eq.(7.5) of ref.\cite{drum:99} contains
the spatial dispersion (in the 'effective mass' approximation), so that $%
n=n\left( \omega ,k\right) $. Inclusion of such a spatial dispersion yields
more than one value of $k$ for certain frequencies \cite{agran:84}, the
additional solutions representing the exciton-like modes characterized by
much larger $k$.

\end{document}